\documentstyle [12pt] {article}
\begin{document}

\pagestyle{empty}
\rightline{\vbox{
\halign{&#\hfil\cr
&hep-ph/9403401 \cr
&FERMILAB-PUB-94/040-T \cr
&NUHEP-TH-94-4 \cr
&UCD-94-3 \cr
&March 1994 \cr}}}
\bigskip
\bigskip
\bigskip
{\Large\bf
	\centerline{Gluon Fragmentation into P-wave Heavy Quarkonium}}
\bigskip
\normalsize

\centerline{Eric Braaten\footnotemark[1]
\footnotetext[1]{on leave from Dept. of Physics and Astronomy,
	Northwestern University, Evanston, IL 60208}}
\centerline{\sl Fermi National Accelerator Laboratory, P.O. Box 500,
Batavia, IL 60510}
\bigskip

\centerline{Tzu Chiang Yuan}
\centerline{\sl Davis Institute for High Energy Physics}
\centerline{\sl Department of Physics, University of California,
    Davis, CA  95616}
\bigskip

\begin{abstract}
The fragmentation functions for gluons to split into P-wave heavy quarkonium
states are calculated to leading order in the QCD coupling constant.
Long-distance effects are factored into two nonperturbative  parameters:
the derivative of the radial wavefunction at the origin and a second
parameter related to the probability  for a heavy-quark-antiquark pair
that is produced in a color-octet S-wave state to form a color-singlet
P-wave bound state.  The fragmentation probabilities for a
high transverse momentum gluon to split into the P-wave
charmonium states $\chi_{c0}$, $\chi_{c1}$, and $\chi_{c2}$ are estimated
to be $0.4 \times 10^{-4}$, $1.8 \times 10^{-4}$, and $2.4 \times 10^{-4}$,
respectively.  This fragmentation process may account for a significant
fraction of the rate for the inclusive production of $\chi_{cJ}$
at large transverse momentum in $p \bar p$ colliders.

\end{abstract}

\vfill\eject\pagestyle{plain}\setcounter{page}{1}

Heavy quarkonium plays an important role in high energy collider
physics, because these states can probe physical processes at short
distances of order $1/m_Q$, where $m_Q$ is the heavy quark mass.
Of particular importance experimentally are the $1^{--}$ S-wave states
of charmonium and bottomonium, which have very clean signatures through their
leptonic decay modes, and the $J^{++}$ P-wave states
with $J = 0,1,2$, which can also be observed through
their radiative transitions into the $1^{--}$ states.
In most previous studies of the direct production of heavy quarkonium
\cite{br},
the dominant production mechanisms were assumed to be given by the Feynman
diagrams that were lowest order in the QCD coupling constant $\alpha_s$.
It has recently been pointed out that the
dominant mechanism for heavy quarkonium production at large
transverse momentum $p_T$ is {\it fragmentation}, the production of a high
energy parton with even larger transverse momentum which subsequently decays
into the quarkonium state plus other partons \cite{by}.
While this mechanism is often of higher order in the QCD coupling
constant $\alpha_s$ than conventional mechanisms, it is
enhanced at large transverse momentum by powers of $p_T/m_Q$,
and thus dominates at sufficiently large $p_T$.

The fragmentation of a parton $i$ into any hadron $H$
is described by a universal fragmentation function $D_{i \to H}(z,\mu)$,
where $z$ is the longitudinal momentum fraction of the
hadron relative to the parton and $\mu$ is a
factorization scale of order $p_T$ \cite{cs}.
If the fragmentation function is known at some initial momentum scale $\mu_0$,
then it can be determined at larger momentum scales $\mu$ by solving the
Altarelli-Parisi evolution equations which sum up the leading
logarithms of $\mu/\mu_0$.
In Ref. \cite{by}, it was shown that in the case of heavy quarkonium,
the fragmentation function $D(z,m_Q)$ at an initial scale of order $m_Q$
can be calculated using perturbative QCD.  The initial fragmentation
functions for gluons to split into S-wave states of heavy quarkonium
were calculated  to leading order in $\alpha_s$ \cite{by}.
The fragmentation functions for heavy quarks to split into S-wave states
have also been calculated  to leading order \cite{bcy1,cc,flsw},
and these calculations have recently been extended to the
P-wave states \cite{py}.

In this paper, we calculate the fragmentation functions for gluons to split
into the P-wave states to leading order in $\alpha_s$. For the sake of clarity,
we describe the calculation in terms of the lowest P-wave states of the
charmonium system:  the $J^{PC} = J^{++}$ states $\chi_{cJ}$,
with $J =0,1,2$, and the $1^{+-}$ state $h_c$.  Our results apply
equally well to the higher P-wave states of charmonium, as well as
to the P-wave states of the bottomonium system.
While a gluon can split into $\chi_{cJ}$ at order
$\alpha_s^2$ through the Feynman diagram in Figure 1, gluon splitting
into the $h_c$ occurs first at one order higher in
$\alpha_s$. Since $1^{+-}$ states of heavy quarkonium like the
$h_c$ are difficult to observe experimentally,
we concentrate in this paper on the $J^{++}$ states.

In charmonium, the charmed quark and antiquark are nonrelativistic
with typical velocity $v \ll 1$ and typical separation $1/(m_c v)$.
Our calculation of the fragmentation function is based on separating
short-distance effects involving the scale $1/m_c$ from long-distance
effects involving scales of order $1/(m_c v)$ or larger.
There are two
distinct mechanisms that contribute to the fragmentation function
at leading order in $v$ \cite{bbl1}, and we will refer to them as the
{\it color-singlet mechanism} and the {\it color-octet mechanism}.
The {\it color-singlet mechanism} is the production of a $c {\bar c}$ pair
in a color-singlet $^3P_J$ state with separation of order
$1/m_c$ in the quarkonium rest frame.  The subsequent formation of the
$\chi_{cJ}$ is a long-distance process with probability of order
$v^5$.  In addition to the volume factor $(m_c v/m_c)^3$, there is an extra
suppression factor of $v^2$ from the wavefunction of the P-state
near the origin.  The {\it color-octet mechanism} is the production of
a $c {\bar c}$ pair in a color-octet $^3S_1$ state with separation of
order $1/m_c$.  The subsequent formation of the $\chi_{cJ}$ can
proceed either through the dominant $|c {\bar c} \rangle$ component
of the $\chi_{cJ}$ wavefunction or through the small $|c {\bar c} g \rangle$
component, which has a probability of order $v^2$.  In the first case,
the $c {\bar c}$ pair must radiate a soft gluon to make a transition to
the color-singlet $^3P_J$ $| c {\bar c} \rangle$ state.
In the second case, a soft gluon must combine with the $c {\bar c}$ pair
to form a color-singlet $| c {\bar c} g \rangle$ state.
In either case, the probability is of order $v^5$, with a volume factor of
$v^3$ and an additional suppression of $v^2$ coming either from the
probability of radiating a soft gluon or from the small probability of the
$|c {\bar c} g \rangle$ component of the wavefunction.
Since the color-singlet mechanism and the color-octet mechanism contribute
to the fragmentation function at the same order in $v$, they must both be
included for a consistent calculation.

Separating effects due to short distances of order $1/m_c$
from those of longer distance scales requires the introduction of an arbitrary
factorization scale $\Lambda$ in the range $m_c v \ll \Lambda \ll m_c$.
The fragmentation functions for heavy quarkonium satisfy factorization
formulas that involve this arbitrary scale.
At leading order in $v^2$, the factorization formula for the
fragmentation function for $g \to \chi_{cJ}$ has two terms:
\begin{equation} {
D_{i \rightarrow {\chi_{cJ}}}(z,m_c) \; = \;
{H_1 \over m_c} d^{(J)}_1(z,\Lambda)
+ (2J+1) {H'_8(\Lambda) \over m_c} d_8(z) \; ,
} \label{fragfact} \end{equation}
where $H_1$ and $H_8'(\Lambda)$ are nonperturbative long-distance factors
associated with the color-singlet and color-octet mechanisms, respectively.
The short-distance factors $d_1^{(J)}(z,\Lambda)$
and $d_8(z)$ can be calculated
using perturbation expansions in $\alpha_s(m_c)$.  They are proportional
to the fragmentation functions for a gluon to split into
a $c {{\bar c}}$ pair with vanishing relative momentum and in the appropriate
color-spin-orbital state: color-singlet $^3P_J$ state for $d_1^{(J)}$ and
color-octet $^3S_1$ state for $d_8$.
Note that in the factorization formula (\ref{fragfact}), the only dependence
on $\Lambda$ is in $d_1^{(J)}$ and $H_8'$.
This simple form holds if the coefficients are calculated
at most to next-to-leading order in $\alpha_s$.  Beyond that order, $d_8$
also acquires a weak dependence on $\Lambda$ and $J$.

The nonperturbative parameters $H_1$ and $H'_8$
can be rigorously defined as matrix elements of 4-quark operators
in nonrelativistic QCD \cite{gpl}.
Their dependence on $\Lambda$ is given by renormalization group equations
whose coefficients can be calculated as perturbation series in
$\alpha_s(\Lambda)$ \cite{bbl1}.
To order $\alpha_s$, $H_1$ is scale invariant and $H_8'$ satisfies
\begin{equation} {
\Lambda {d \ \over d \Lambda} H'_8(\Lambda)
\; = \; {16 \over 27\pi}
	\alpha_s(\Lambda)H_1 \; .
} \label{rgeq} \end{equation}
If the factorization scale $\Lambda$ is chosen to be much less than $m_c$,
this equation can be used to sum up large logarithms of $m_c/\Lambda$:
\begin{equation} {
H'_8(m_c) \;  = \; H'_8(\Lambda)
\; + \; {16 \over 27 \beta_0}
	\log \left( {\alpha_s(\Lambda) \over \alpha_s(m_c)} \right) \; H_1 \; ,
} \label{rgsol} \end{equation}
where $\beta_0 = (33 - 2 n_f)/6$ is the first coefficient in the beta
function for QCD with $n_f$ flavors of light quarks.
The parameter $H_1$ can be related to  the derivative of the
nonrelativistic radial wavefunction at the origin for the P-wave states:
\begin{equation} {
H_1 \; \approx \; {9 \over 2\pi}  {|R'_P(0)|^2 \over m_c^4} \;
\left( 1 + {\cal O}(v^2) \right) \;.
} \label{hone} \end{equation}
This parameter can be determined phenomenologically from the
annihilation rates of the $\chi_{cJ}$ states.
Using recent high precision measurements of the light hadronic decay
rates of $\chi_{c1}$ and $\chi_{c2}$, $H_1$ has been determined to be
approximately ${\rm 15 \; MeV}$ \cite{bbl2}.  The parameter $H_8'$ was
introduced in Ref. \cite{bbly} in a calculation of the rate for the decay
$b \to \chi_{cJ}+X$, which also receives contributions from both the
color-singlet and color-octet mechanisms for $\chi_{cJ}$ production.
The prime on $H_8'$ is a reminder that
this parameter is not related in any rigorous way to
the corresponding parameter $H_8$ that appears in decays of the $\chi_{cJ}$
states into light hadrons.  Using data on the inclusive decays of
$B$-mesons into charmonium, its value for $\Lambda = m_c$
has been estimated to be $H_8'(m_c) \approx 3 \; {\rm MeV}$ \cite{bbly}.
This parameter also enters into the inclusive decay rate
of the $\Upsilon$ into P-wave charmonium states \cite{ht}.

We now turn to the calculation of the coefficient $d_1^{(J)}(z,\Lambda)$
in the color-singlet contribution to the fragmentation function.
We follow the method and notation of Ref. \cite{by}.
Let ${\cal A}_\alpha$ denote the amplitude
for $g^* \rightarrow c {\bar c} (^3P_1,{\underline 1}) + g$
corresponding to the Feynman diagram in Figure 1.
The $c \bar c$ pair have almost equal momenta, and are in a
color-singlet $^3P_J$ state.
The amplitude ${\cal A}_\alpha$ can be written down in terms of $R_P'(0)$
using standard Feynman rules for quarkonium processes \cite{kks}.
Multiplying ${\cal A}_\alpha$ by its complex conjugate and summing over
final colors and spins, we obtain the following generic form
\begin{equation} {
\sum {\cal A}_\alpha {\cal A}_\beta^* \;
= \; {H_1 \over m_c} \; \Bigg(
A_J(s) (-g_{\alpha \beta})
+ B_J(s) p_\alpha p_\beta
+ C_J(s) (p_\alpha q_\beta + q_\alpha p_\beta)
+ D_J(s)  q_\alpha q_\beta  \Bigg) \;,
} \label{Asq} \end{equation}
where $p$ and $q$ are the 4-momenta of the $c \bar c$ pair and
the fragmenting gluon $g^*$, respectively, and $s = q^2$.
The strategy is to reduce this expression in the limit $q_0 \gg m_c$
to the polarization sum $(-g _{\alpha \beta} + ...)$ for an on-shell gluon
multiplied by a function of $s$ and $z$, where $z$ is the longitudinal
momentum fraction of the $c \bar c$ pair relative to the fragmenting gluon.
Terms  in (\ref{Asq}) that are proportional to $q_\alpha$ or $q_\beta$
can be dropped, because in axial gauge, $q_\alpha$ and $q_\beta$
are of order $m_c^2/q_0$ when contracted with the numerator of the
propagator of the virtual gluon.
In the $p_\alpha p_\beta$ term, we can set $p = zq + p_\perp$,
where $p_\perp$ is the transverse part of the 4-vector $p$.
After averaging over the directions of $p_\perp$,
$p_\alpha p_\beta$ can be replaced by
$-g_{\alpha \beta} {{\vec p}_\perp}^{\;2}/2$, up to terms that are suppressed
in axial gauge.
The terms in (\ref{Asq}) that contribute to fragmentation then reduce to
\begin{equation} {
\sum {\cal A}_\alpha {\cal A}_\beta^*
\; \approx \;  {H_1 \over m_c} \;
\left( A_J(s) + {{{\vec p}_\perp}^{\;2} \over 2} B_J(s) \right) \;
	(- g_{\alpha \beta})  \;.
} \label{Asqfrag} \end{equation}
Energy-momentum conservation in the form
$s = ({\vec p}_\perp^{\;2} + 4m_c^2)/z + {\vec p}_\perp^{\;2}/(1-z)$
can be used to eliminate ${\vec p}_\perp^{\;2}$ in (\ref{Asqfrag})
in favor of $s$ and $z$.
The fragmentation probability is obtained by dividing
the coefficient of $(- g_{\alpha \beta})$ by $s^2$ for the propagator of the
virtual gluon, and then integrating over the
phase space of the $c \bar c$ pair and the gluon in the final state.
The phase space integral can be expressed compactly in terms of
integrals over $s$ and $z$ \cite{by}.
The resulting expression for the integral over $z$ of
$d_1^{(J)}(z,\Lambda)$ is
\begin{equation} {
\int_0^1 dz \; d^{(J)}_1(z,\Lambda)
\; = \; {1 \over 16 \pi^2} \int^\infty_{s_{\rm min}(\Lambda)} ds
\int_{4m_c^2/s}^1 dz \; {1 \over s^2} \left( A_J(s)
	\; + \; {(1-z)(z s - 4 m_c^2) \over 2} B_J(s) \right) \; .
} \label{fragp} \end{equation}
We have anticipated the presence of an infrared divergence
associated with a soft gluon in the final state by imposing a lower cutoff
$\Lambda$ on the energy of the gluon in the quarkonium rest frame.
This translates into a lower limit on $s$:
$s_{\rm min}(\Lambda) = 4 m_c^2(1 + \Lambda/m_c)$.
The calculations of the functions $A_J(s)$ and $B_J(s)$ in (\ref{fragp})
involves some rather complicated algebra, but the final results are
relatively simple.  Interchanging orders of integration in (\ref{fragp}),
we can read off the functions $d_1^{(J)}(z,\Lambda)$:
\begin{eqnarray}
d_1^{(J)}(z,\Lambda)
&=& {\alpha_s^2 \over 27} \int^\infty_{4 m_c^2/z} ds
\; {m_c^2 \over s^2(s-4m_c^2)^4} \; f_J(s,z) \; ,
\quad z < \left( 1 \;+\; {\Lambda \over m_c} \right)^{-1} \;,
\label{fragprobz} \\
&=& {\alpha_s^2 \over 27} \int^\infty_{s_{\rm min}(\Lambda)} ds
\; {m_c^2 \over s^2(s-4m_c^2)^4} \; f_J(s,z) \; ,
\quad z > \left( 1 \;+\; {\Lambda \over m_c} \right)^{-1} \;,
\label{fragprob1} \end{eqnarray}
where
\begin{eqnarray}
f_0(s,z) &=& (s - 12 m_c^2)^2
\left[ (s - 4 m_c^2)^2 \;-\; 2 (1-z) (z s - 4 m_c^2) s \right] \; ,
\label{f0} \\
f_1(s,z) &=& 6 s^2
\left[ (s - 4 m_c^2)^2  \;-\; 2 (1-z) (z s - 4 m_c^2) (s - 8 m_c^2) \right] \;
,
\label{f1} \\
f_2(s,z) &=& 2
\left[ (s - 4 m_c^2)^2 (s^2 + 96 m_c^4) \;-\; 2 (1-z) (z s - 4 m_c^2)
	s (s^2 - 24 s m_c^2 + 96 m_c^4) \right] \; .
\label{f2} \end{eqnarray}
For $z < (1+\Lambda/m_c)^{-1}$, the integral over $s$ in (\ref{fragprobz})
can be calculated straightforwardly.  The cutoff $\Lambda$ can be set to
zero everywhere except in terms proportional to $1/(1-z)$, which diverge
upon integrating over $z$. In the $1/(1-z)$ terms, the limit
$\Lambda \ll m_c$ must be taken more carefully, and it gives rise to
a plus-distribution:
\begin{equation} {
{1 \over 1-z} \; \theta \left( 1 - z - {\Lambda \over m_c} \right) \;
\longrightarrow \; {1 \over (1-z)_+} \; - \;
\log {\Lambda \over m_c} \; \delta(1 - z) \; .
} \label{plusf} \end{equation}
For $z > (1+\Lambda/m_c)^{-1}$, the limit $\Lambda \ll m_c$ can be taken
only after evaluating the integral over $s$ in (\ref{fragprob1}).  This
gives rise to additional endpoint contributions proportional to
$\delta(1-z)$.  Our final result for the short distance factor multiplying
$H_1/m_c$ in the fragmentation function is
\begin{equation} {
d_1^{(J)}(z,\Lambda) \;=\; {2 \alpha_s^2 \over 81}
\left[ (2J+1) {z \over (1-z)_+}
	\;+\; \left( Q_J - (2J+1) \log{\Lambda \over m_c} \right) \delta(1-z)
	\; + \; P_J(z) \right] \; ,
} \label{done} \end{equation}
where the coefficients $Q_J$ are
\begin{equation} {
Q_0 \;=\; {13 \over 12} \;, \quad
Q_1 \;=\; {23 \over 8} \;, \quad
Q_2 \;=\; {121 \over 24} \;, \quad
} \label{qJ} \end{equation}
and the functions $P_J(z)$ are
\begin{eqnarray}
P_0(z) &=& {z (85 -  26 z) \over 8}
	\; + \; {9 (5 - 3z) \over 4} \; \log(1-z) \; ,
\label{pzero} \\
P_1(z) &=& - {3 z (1 + 4 z) \over 4} \; ,
\label{pone} \\
P_2(z) &=& {5 z (11 - 4z) \over 4}
	\; + \; 9 (2-z) \; \log(1-z) \; .
\label{ptwo} \end{eqnarray}

We next consider the color-octet coefficient $d_8(z)$ in the
fragmentation formula (\ref{fragfact}).  At leading order in $\alpha_s$,
this contribution to the fragmentation function comes
from the subprocess $g^* \to c {\bar c}(^3S_1, {\underline 8})$
given by the Feynman diagram
in Figure 2.  The $c$ and ${\bar c}$ have equal momenta $q/2$,
and are in a color-octet $^3S_1$ state.  The projection onto this state
can be reduced to a simple Feynman rule:
\begin{equation} {
v(q/2) {\bar u}(q/2) \; \to \;
{R_8(0) \over \sqrt{16 \pi m_c}} T^a_{ij} \not\!\epsilon  (q)
(\not\! q  + 2 m_c) \; ,
} \label{rule} \end{equation}
where $\epsilon^\mu(q)$ is the polarization 4-vector of the $^3S_1$ state
and $i$, $j$, and $a$ are the color indices of the quark, antiquark,
and color-octet state, respectively.  The parameter $R_8(0)$ is a
fictitious ``color-octet radial wavefunction at the origin'' related to the
nonperturbative matrix element $H_8'(\Lambda)$ by
$H_8' = (2/3 \pi) |R_8(0)|^2/m_c^2$.  The square of the amplitude
${\cal A}_\alpha$ for the subprocess $g^* \to c {\bar c}$,
summed over final state colors and spins, is
\begin{equation} {
\sum {\cal A}_\alpha {\cal A}_\beta^* \;
= \; 6 \pi \alpha_s m_c^3 H_8'(\Lambda) \;
\left ( - g_{\alpha \beta} + {q_\alpha q_\beta \over 4 m_c^2} \right) \; .
} \label{Asqeight} \end{equation}
The $q_\alpha q_\beta$ term can be dropped because $q_\alpha$ is of order
$m_c^2/q_0$ when contracted with the numerator of the virtual gluon
propagator in axial gauge.  The expression therefore reduces to the
polarization sum $(- g_{\alpha \beta} + ...)$ for an on-shell gluon
multiplied by $ 6 \pi \alpha_s m_c^3 H_8'$.  Dividing by $(4 m_c^2)^2$
for the virtual gluon propagator, we obtain
the fragmentation probability $(3 \pi \alpha_s/8) H_8'/m_c$.
This probability can be identified with the second term in
(\ref{fragfact}), integrated over $z$ and summed over $J=0,1,2$.
This term in the fragmentation function contributes only
at the endpoint $z = 1$.  We can therefore identify the function
$d_8(z)$ in (\ref{fragfact}) to be
\begin{equation} {
d_8(z) \; = \;
{\pi \alpha_s \over 24} \; \delta (1-z) \;.
} \label{deight} \end{equation}

The total fragmentation function at leading order in $\alpha_s$ is
given by the factorization formula (\ref{fragfact}), with the color-singlet
coefficient given in (\ref{done}) and the color-octet coefficient
given in (\ref{deight}).  To avoid large logarithms of $m_c /\Lambda$
in the color-singlet coefficient, we can choose $\Lambda = m_c$.
We thus arrive at the final expressions for the fragmentation functions
of gluon splitting into $\chi_{cJ}$ to leading order in $\alpha_s$:
\begin{eqnarray}
D_{g \rightarrow \chi_{cJ}}(z,2m_c)
\; & \approx & \; {2 \alpha_s^2(2m_c) \over 81}{H_1 \over m_c}
\Bigg[ (2J+1) {z \over (1-z)_+}
	\; + \; Q_J \; \delta(1-z) \; + \; P_J(z) \Bigg]
\nonumber \\
&& + \; (2J+1) \; {\pi\alpha_s(2m_c) \over 24}{H'_8(m_c) \over m_c} \;
\delta (1-z)  \; ,
\label{Dfinal} \end{eqnarray}
where $Q_J$ and $P_J(z)$ given by (\ref{qJ})-(\ref{ptwo}).
The choice of the scale $\mu$ in the running coupling constant is
independent of the choice of factorization scale $\Lambda$.
We have followed Ref. \cite{by} in choosing $\mu = 2 m_c$,
which is the minimum value of the invariant mass of the virtual gluon.
If we wish to use a value for the
factorization scale $\Lambda$ in (\ref{Dfinal}) which is significantly
smaller than $m_c$, we should use the solution (\ref{rgsol}) to the
renormalization group equation for $H_8'(\Lambda)$
to sum up the leading logarithms of $m_c / \Lambda$.

Rough  estimates of the gluon fragmentation contribution
to  the production of the $\chi_{cJ}$ states at large transverse momentum
in any high energy process can be obtained by multiplying
the cross sections for producing gluons with transverse momentum
larger than $2 m_c$ by appropriate fragmentation probabilities.
Integrating the initial fragmentation functions (\ref{Dfinal}) over $z$,
we obtain the probabilities
\begin{equation} {
P_{g \rightarrow \chi_{cJ}} \; \approx \;
- R_J \;  {\alpha_s^2(2m_c) H_1 \over 108 m_c} \; + \;
(2J+1) \; {\pi \alpha_s(2m_c) H'_8(m_c) \over 24 m_c} \; ,
} \label{PchiJ} \end{equation}
where $R_0 = 5$, $R_1 = 4$, and $R_2 = 16$.
Notice that with the choice $\Lambda = m_c$ for the factorization scale,
the color-singlet pieces give rise to negative contributions
to the initial fragmentation probabilities. Requiring that all the
probabilities (\ref{PchiJ}) be positive, we obtain an interesting lower
bound on $H'_8(m_c)$:
\begin{equation} {
H'_8(m_c) \; > \; {10 \alpha_s(2m_c) \over 9 \pi} \; H_1  \; .
} \label{heightlowerbound} \end{equation}
Using $H_1 \approx 15 \; {\rm MeV}$,  $m_c = 1.5 \; {\rm GeV}$,
and $\alpha_s(2m_c) = 0.26$, we find $H'_8(m_c) \; > \; 1.4 \; {\rm MeV}$.
The estimate $H'_8(m_c) \approx 3 \; {\rm MeV}$ obtained in Ref. \cite{bbly}
is consistent with this lower bound. Using the value
$H'_8(m_c) \approx 3 \; {\rm MeV}$, our estimates for the
the initial fragmentation probabilities in (\ref{PchiJ})
are $0.4 \cdot  10^{-4}$, $1.8 \cdot 10^{-4}$, and
$2.4 \cdot 10^{-4}$ for $\chi_{c0}$, $\chi_{c1}$, and $\chi_{c2}$,
respectively. The production of $\chi_{cJ}$ states contributes to the
inclusive rate for production of the $1^{--}$ charmonium state
$J/\psi$ through the radiative decay $\chi_{cJ} \to J/\psi + \gamma$.
Multiplying the fragmentation probabilities given above by the
appropriate radiative branching fractions of $0.7 \%$, $27 \%$, and $14 \%$,
we find that the probability of a $J/\psi$ in a gluon jet is
approximately $8 \times 10^{-5}$. This is more than an order of
magnitude larger than the probability $3 \times 10^{-6}$ for the direct
fragmentation of a gluon into $J/\psi$ that was obtained in Ref. \cite{by}.

The methods used above to calculate the fragmentation functions
$D_{g \to \chi_{cJ}}(z)$ can also be used to calculate
the distribution of the transverse momentum $p_\perp$ of the $\chi_{cJ}$
relative to the gluon jet. This transverse momentum is related to the
the invariant mass $s$ of the gluon jet by
$s = (\vec p_\perp^{\,2} + 4m_c^2)/z + \vec p_\perp^{\,2}/(1-z)$.
For the color-singlet
contribution, the $s$-distribution is obtained by integrating over
$z$ in (\ref{fragp}).  For the color-octet contribution, the
$s$-distribution is a delta-function at $s = 4 m_c^2$.  Adding these
two contributions we obtain
\begin{eqnarray}
{d P_{g \rightarrow \chi_{c0}} \over ds}
&=& {2 \alpha_s^2 m_c H_1 \over 81} \;
	{(s-12m_c^2)^2 \over s^3 (s-4m_c^2)} \;
	\theta(s - s_{\rm min}(\Lambda))
\; + \; {\pi \alpha_s H_8'(\Lambda) \over 24 m_c}
\; \delta(s - 4 m_c^2) \; ,
\label{probzero} \\
{d P_{g \rightarrow \chi_{c1}} \over ds}
&=& { 4\alpha_s^2 m_c H_1 \over 27} \;
	{s+4m_c^2 \over s^2(s-4m_c^2)} \; \theta(s - s_{\rm min}(\Lambda))
\; + \; 3 {\pi \alpha_s H_8'(\Lambda) \over 24 m_c}
\; \delta(s - 4 m_c^2) \; ,
\label{probone} \\
\label{probtwo}
{d P_{g \rightarrow \chi_{c2}} \over ds}
&=& {4 \alpha_s^2 m_c H_1 \over 81} \;
	{s^2+12sm_c^2+96m_c^4 \over s^3(s-4m_c^2)}
	\; \theta(s - s_{\rm min}(\Lambda))
\; + \; 5 {\pi \alpha_s H_8'(\Lambda) \over 24 m_c}
\; \delta(s - 4 m_c^2) \; , \nonumber \\
&& \;
\end{eqnarray}
where $s_{\rm min}(\Lambda) = 4 m_c^2 (1 + \Lambda/m_c)$.
Integrating over $s$, we recover the fragmentation probabilities
given in (\ref{PchiJ}).  The cutoff-dependence of the color-singlet terms in
(\ref{probzero})-(\ref{probtwo}) is cancelled by the $\Lambda$-dependence
of the parameter $H_8'(\Lambda)$ in the color-octet terms.
The color-singlet terms in these invariant mass distributions were
obtained previously by Hagiwara, Martin and Stirling \cite{hms},
up to an error of $4\pi$ in the overall coefficient.
They did not include the color-octet contributions,
so their answers were sensitive to the
value of the infrared cutoff $\Lambda$.
In the region near the lower endpoint $s = 4 m_c^2$,
the distributions (\ref{probzero})-(\ref{probtwo}) must of course be
smeared over an appropriate range in $p_\perp$ before they can be
compared with experimental data.

We have calculated the fragmentation functions for gluons to
split into P-wave quarkonium states to leading
order in $\alpha_s$.  The fragmentation functions satisfy a factorization
formula with two nonperturbative parameters $H_1$ and $H'_8$
which can be determined from other processes involving the
annihilation and production of P-wave states.
These fragmentation functions are universal and can be used to calculate the
rates for the direct production of P-wave states at large transverse
momentum in any high energy process.  They are also needed
to calculate the total production rate of the $1^{--}$ states
from the fragmentation mechanism, since the P-wave states
have significant rates for transitions to the $1^{--}$ states.
The fragmentation probabilities for $g \to \chi_{c1}$ and $g \to \chi_{c2}$
were estimated to be on the order of $10^{-4}$.
This is large enough that gluon fragmentation into $\chi_{cJ}$ should
account for a significant fraction of the $\chi_{cJ}$'s that are
observed at large transverse momentum in hadron colliders.
Fragmentation into $\chi_{cJ}$ followed by its radiative decay
may also account for a significant fraction of the $J/\psi$'s
that are produced at large $p_T$.

While this paper was being written, we received a paper by Ma \cite{ma},
in which the color-singlet term in the fragmentation function
for $g \to \chi_{c1}$ is calculated for longitudinally and transversely
polarized $\chi_{c1}$ separately. After summing over polarizations,
his result agrees with ours except near the endpoint $z = 1$.
In the endpoint region, Ma's fragmentation function is sensitive to an
infrared cutoff.

This work was supported in part by the U.S. Department of Energy,
Division of High Energy Physics, under Grants DE-FG02-91-ER40684 and
DE-FG03-91ER40674 and also by the Texas National Research Laboratory
Commission Grant RGFY93-330.

\bigskip

\noindent{\Large\bf Figure Captions}
\begin{enumerate}
\item Feynman diagrams 	for $g^* \rightarrow c {\bar c} + g$ which
	contribute to the color-singlet term in the fragmentation function
	for $g \to \chi_{cJ}$.
\item Feynman diagram 	for $g^* \rightarrow c {\bar c}$ which
	contributes to the color-octet term in the fragmentation function
	for $g \to \chi_{cJ}$.
\end{enumerate}
\vfill\eject

\end{document}